\documentclass[5p,twocolumn,times]{elsarticle}
\journal{Physics Letters B}

\usepackage{graphicx}  % include figure files
\usepackage{amsmath}
\usepackage{amssymb}
\usepackage{tikz}
\usepackage{float}
\usepackage{epstopdf}
\usepackage{dcolumn}% Align table columns on decimal point
\usepackage[percent]{overpic}  % insert LaTex text over figures
\usepackage{bm}
\usepackage{xcolor,xspace}
\usepackage{braket}
\usepackage{wasysym}  % for the use of \CIRCLE
\usepackage[utf8]{inputenc}
\DeclareUnicodeCharacter{2009}{\,}
\usepackage[ulem=normalem]{changes}
\usepackage{hyperref}
\hypersetup{colorlinks=True,urlcolor=blue,linkcolor=blue,citecolor=blue,filecolor=black}

\colorlet{Changes@Color}{red}  % changes in red color

\newcommand{\ba}{\begin{eqnarray}}
\newcommand{\ea}{\end{eqnarray}}
\newcommand{\bsub}{\begin{subequations}}
\newcommand{\esub}{\end{subequations}}

\def\ket#1{|#1\rangle}
\def\bra#1{\langle#1|}
\def\b0{\beta_0}
\def\beq{\beta_{\rm eq}}
\def\g0{\gamma_0}
\def\gaeq{\gamma_{\rm eq}}
\def\bs{\beta_*}

\def\Gaeq{\Gamma_{\rm eq}}

\begin{document}

\title{Rigid triaxial roto-vibrator in a finite system}

\author{A. Leviatan}
\ead{ami@phys.huji.ac.il}


\address{Racah Institute of Physics,
  The Hebrew University, Jerusalem 91904, Israel}

\date{\today}

\begin{abstract}
  We investigate the emergence of rigid triaxiality
  in a finite system, within
  the framework of the interacting boson model of nuclei.
  Appropriate Hamiltonians for
  a maximally-triaxial rigid shape are constructed.
  Closed expressions for wave functions, energies and
  E2 rates for states of the ground band are derived.
  Classical and quantum analyses reveal that for
  a fixed Hamiltonian, the system evolves from
  a $\gamma$-soft SO(6) structure
  for small $N$, to a $\gamma$-rigid structure
  for large $N$. Attention is paid to
  symmetry and spectral properties of both
  rotational and vibrational states.
\end{abstract}


\maketitle

Unraveling the nature (soft vs. rigid) and origin of
triaxiality at low spin in nuclei, is a long standing
challenge. Recent experimental and theoretical developments
have spurred renewed interest in this topic.
Coulomb excitation experiments provided evidence for
rigid triaxiality in
$^{76}$Ge~\cite{Toh13,Ayang19,Ayang23}
and ultra-relativistic heavy-ion collisions
emerged as
a promising tool to probe triaxial deformation
in nuclei~\cite{Bally22,Jia22,Zhao24}.
On the theoretical side, large-scale
Monte Carlo shell model calculations of $^{166}$Er have
questioned its traditional interpretation as axially
deformed, replacing it with triaxial
rigidity~\cite{Tsunoda21} and leading
to claims for prevailing triaxial shapes in
nuclei~\cite{Otsuka25,Bonat25}.

Nuclear triaxiality is traditionally described using two
simple models within the framework of the Bohr
Hamiltonian~\cite{BM}. The rigid triaxial model of
Davydov and Filippov (DF)~\cite{Davydov58,DavydovRost59},
has a well-defined potential minimum
at a nonzero value of $\gamma$,
whereas the $\gamma$-soft
model of Wilets and Jean (WJ)~\cite{Wilets56},
incorporates
a $\gamma$-independent potential. Nuclei are mesoscopic
systems, with a finite number of constituents, hence it is
important to understand how these two paradigms of
triaxiality emerge in such environment.
In the present Letter,
we address this issue
in the framework of the interacting boson model
(IBM)~\cite{ibm}, which
describes quadrupole collective states in even-even
nuclei in terms of $N$ monopole ($s$) and quadrupole ($d$)
bosons, representing valence nucleon pairs.
The model is based on a
U(6) spectrum generating
algebra and encompasses several symmetry limits:
U(5) relevant for spherical,
SU(3) and SO(6) relevant for deformed shapes.

Accommodating
rigid triaxial shapes
in the IBM framework,
necessitates higher-order terms
in the Hamiltonian~\cite{IsaChen81}.
Two approaches have been
implemented for their selection.
A~symmetry-conserving approach considers an expansion
of the Hamiltonian in the generators
of SU(3)~\cite{Smirnov00,Zhang14}
or SO(6)~\cite{Thiamova10,Teng24}.
A second approach
involves a broken-symmetry Hamiltonian obtained
by adding a specific
cubic term to a standard one+two body
Hamiltonian~\cite{HeydeIsa84,Casten85,SorgPVI08}.
In what follows,
we present an alternative approach for which the
Hamiltonian breaks a prescribed symmetry
but maintains a good
quantum number to selected states.
This allows for closed $N$-dependent expressions 
of wave functions and observables which shed light
on the emergence of triaxiality in a finite
system.

Focusing on a maximally-triaxial shape,
in an adiabatic model based on
the rigid-triaxial DF Hamiltonian
with $\gamma\!=\!\pi/6$
(the Meyer-ter-Vehn limit~\cite{Meyer75})
and vibrational terms, the spectrum reads
$E= B[\,4L(L+1) -3K^2\,]
  + (n_{\beta}+1/2)\omega_{\beta}
  + (n_{\gamma}+1/2)\omega_{\gamma}$.
  Here $K$ is the projection of the angular momentum ($L$)
  on the 1-axis and $(n_{\beta},\,n_{\gamma})$ are vibrational
  quantum numbers.
  The ground, one $\beta$-phonon and one $\gamma$-phonon
bands correspond to
$(n_{\beta},n_{\gamma})=(0,0),\,(1,0),\,(0,1)$.
The rotational states $L=0,2^2,3,4^3,5^2,6^4,\ldots$
in each band are arranged in $K$-sequences
with $L=0,2,4,6,\ldots$ for $K\!=\!0$ and
$L=K,K\!+\!1,K\!+\!2,\ldots$~for $K\!>\!0$ even.
The states have good ${\cal R}$-parity given by
${\cal R} = (-1)^{n_{\gamma}+K/2}$~\cite{RoweWood,Rowe09}.
E2 transitions between states with the same
${\cal R}$-parity are forbidden and
all quadrupole moments vanish.

\begin{figure}[t]
\vspace{-1.4cm}
\hspace*{-1.15cm}
\begin{overpic}[width=1.4\linewidth, trim=10mm 3mm 0mm 
    0mm, clip]{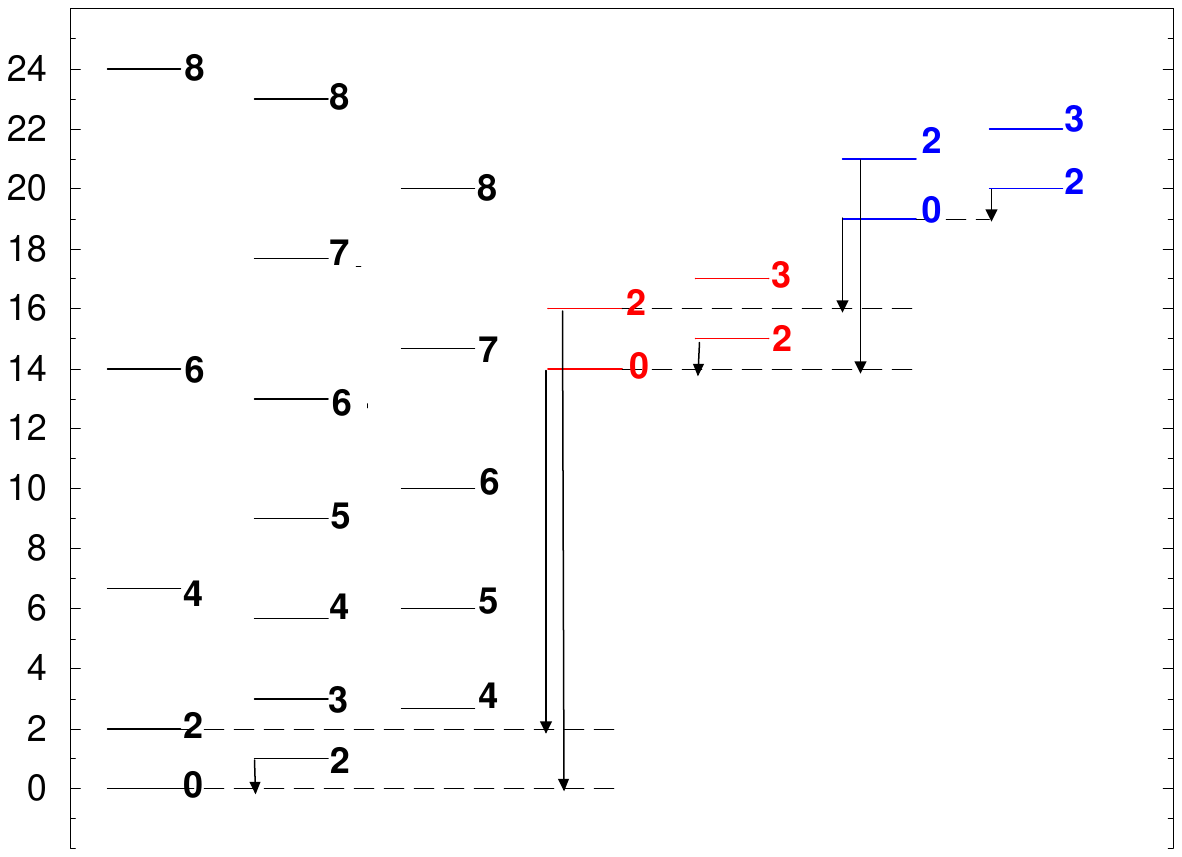}
\put(20,8) {\small ground band}
\put(20,5) {\small $\sigma\!=\!N$}
\put(42,8) {\small \textcolor{red}{$\gamma$ band}}
\put(33,5) {\small \textcolor{red}
  {$\sigma\!=\!N,N\!-\!2,N\!-\!4,N\!-\!6$}}
\put(66,8) {\small \textcolor{blue}{$\beta$ band}}
\put(66,5) {\small \textcolor{blue}
  {$\sigma\!=\!N\!-\!2$}}
\end{overpic}
  \vspace{-1cm}
  \caption{\label{fig-spectra}
  \small
  Typical spectrum of the Hamiltonian, Eq.~(\ref{Hfull})
  in arbitrary units.
  The SO(6) $\sigma$-assignments of the lowest bands,
  Eq.~(\ref{wf-gbgam}), are indicated.
  Each band
  contains $K$-sequences ($K=0,2,4,\ldots$) of rotational
  states.
  Shown are allowed E2 transitions connecting
  $0^{+}$ and $2^{+}$ states.
  }
\end{figure}

\begin{figure}[t]
  \includegraphics[width=0.87\linewidth]
                  {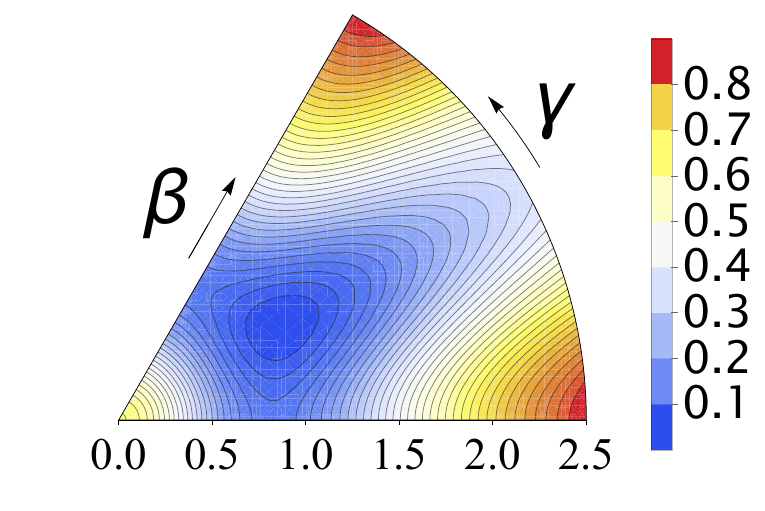}
  \caption{\label{fig-Esurface}
  \small
  Energy surface $\tilde{E}(\beta,\gamma)$,
  Eq.~(\ref{E-surface}),
  for the Hamiltonian~(\ref{Hfull}),
  with parameters
  $h_0\!=\!15,\,r_0\!=\!0.25,\,\rho_5\!=\!\rho_3=0,\,
  N\!=\!25$.
  $\tilde{E}(\beta,\gamma)$ has
  a~global minimum at $(\beta\!=\!1,\gamma\!=\!\pi/6)$,
  a~local maximum at $\beta\!=\!0$ and two
  prolate and oblate saddle~points.
  }
\end{figure}

To study finite-$N$ effects on the dynamics of
such a triaxial shape,
we consider the following IBM Hamiltonian,
\ba
\hspace{-0.5cm}
\hat{H} =
h_0\,P^{\dag}_0P_0
+ r_0\,W^{\dag}_{0}W_0 
+ \rho_5\,\hat{C}_{\rm SO(5)}
+ \rho_3\,\hat{C}_{\rm SO(3)} ~.
\label{Hfull}
\ea
Here $P^{\dag}_{0} = 
d^{\dag}\cdot d^{\dag} - (s^{\dag})^2$,
$W^{\dag}_{0} = 
\sqrt{14}\,(d^{\dag} d^{\dag})^{(2)}\cdot d^{\dag}$ and
$P^{\dag}_0P_0
= -\hat{C}_{\rm SO(6)} + \hat{N}(\hat{N}+4)$,
where $\hat{C}_{\rm G}$ denotes
the Casimir operator
of ${\rm G= SO(6),\,SO(5),\,SO(3)}$
and $\hat{N}=\hat{n}_s+\hat{n}_d = s^{\dag}s
  + \sum_{m}d^{\dag}_md_m$.
A~typical spectrum, shown in Fig.~\ref{fig-spectra},
resembles that of the adiabatic DF-based model
for a triaxial
shape with $\gamma\!=\!\pi/6$. 
This interpretation is consistent with the geometric
visualization of the IBM obtained by
an energy surface,
$E_{N}(\beta,\gamma)$,
defined by the expectation value of the
Hamiltonian
in a coherent (intrinsic) state~\cite{GinoKir80,Diep80},
\ba
\ket{\beta,\gamma;N} = (N!)^{-1/2}(b^\dag_c)^N\ket{0} ~,
\label{int-state}
\ea
where
$b^\dag_c \!=\! \tfrac{1}{\sqrt{1+\beta^2}}
[\beta\cos\gamma d^\dag_0 +
  \tfrac{1}{\sqrt{2}}\beta\sin\gamma
  (d^\dag_2 + d^\dag_{-2}) + s^\dag]$.
Here $(\beta,\gamma)$ are quadrupole shape parameters
whose values,
$(\beta_{\rm eq},\gamma_{\rm eq})$, 
at the global minimum of $E_{N}(\beta,\gamma)$ define
the equilibrium shape
and $\vert\beq,\gaeq ; N \rangle$ serves as the
intrinsic state
for the ground band.
For $\hat{H}$
of Eq.~(\ref{Hfull}) we find
$E_{N}(\beta,\gamma) \!=\!
N(N-1)(N-2)\tilde{E}(\beta,\gamma)$,
\ba
\tilde{E}(\beta,\gamma) =
\bar{h}_{0}\frac{(\beta^2 - 1)^2}{(1+\beta^2)^{2}}
+ 4r_0\frac{\beta^6}{(1+\beta^2)^{3}}\Gamma^2 ~,
\label{E-surface}
\ea
where $\Gamma\equiv \cos 3\gamma$ and
$\bar{h}_0 = h_0/(N-2)$.
The contribution of the $\rho_5$ and $\rho_3$ terms in
Eq.~(\ref{Hfull}) to the energy surface is $1/N^2$
suppressed, hence negligible
for large $N$.
$\tilde{E}(\beta,\gamma)$ has a global triaxial
minimum,
\ba
\beq=1\;,\;\Gaeq=0 \;\; (\gaeq=\pi/6) ~,
\label{b1gampi6}
\ea
at $\tilde{E}=0$.
The surface has also a
spherical ($\beta=0$) local maximum at
$\tilde{E}=\bar{h}_0$
and two degenerate prolate and oblate saddle points,
$[\bs =
  (1+3r_0/\bar{h}_0)^{-1/4},\,
  \gamma_{*}=0,\,\tfrac{\pi}{3}]$,
at $\tilde{E}= \tfrac{1}{3}\bar{h}_0(1+\bs^2)^{-2}
(1-\bs^2)(3+\bs^2)$.
A typical energy surface landscape is shown in
Fig.~\ref{fig-Esurface}.
It is convenient to use
$\gaeq=\pi/2$
instead of
$\gaeq=\pi/6$
in Eq.~(\ref{b1gampi6}).
Both values of $\gamma$ give the same shape
since the
two intrinsic states are related
by a rotation
which carries the 1-axis to the 3-axis.

Normal modes correspond to small oscillations about the
global minimum of the energy surface. For the
triaxial shape of
Eq.~(\ref{b1gampi6}), there are two such modes
involving $\beta$ and $\gamma$ vibrations.
For large $N$, their frequencies
read~\cite{LevShao90},
\bsub
\ba
\epsilon_{\beta} &=&
4N^2\, \bar{h}_0  ~,\\
\epsilon_{\gamma} &=&
9N^2\,r_0 ~.
\ea
\label{ebet-egam}
\esub
The corresponding modes are represented by 
$b^{\dag}_{\beta} = \tfrac{1}{\sqrt{1+\beta^2}}
[\cos\gamma\,
  d^{\dag}_{0} + \tfrac{1}{\sqrt{2}}\sin{\gamma} 
  (d^{\dag}_{2} + d^{\dag}_{-2}) - \beta\, s^{\dag}]$
and
$b^{\dag}_{\gamma} =
\tfrac{1}{\sqrt{2}}\cos{\gamma}\,
( d^{\dagger}_{2} + d^{\dagger}_{-2})
-\sin\gamma\,
d^{\dagger}_{0}$
evaluated at
$(\beta\!=\!\beq,\gamma\!=\!\gaeq)$.
Both modes are one-dimensional and represent
single-phonon excitations of a triaxial shape.
The ratio
$\epsilon_{\gamma}/\epsilon_{\beta}=
\tfrac{9}{4}(r_0/\bar{h}_0)$ determines the relative
position of the~bands.

The Hamiltonian of Eq.~(\ref{Hfull}) naturally
splits into intrinsic and collective parts,
$\hat{H} = \hat{H}_{\rm int}(h_0,r_0) +
\hat{H}_{\rm col}(\rho_5,\rho_3)$.
The intrinsic Hamiltonian, $\hat{H}_{\rm int}$,
is constructed to satisfy
$\hat{H}_{\rm int}\ket{\beq\!=\!1,
  \gaeq\!=\!\tfrac{\pi}{2};N}=0$.
It determines the energy surface and band structure.
The collective Hamiltonian, $\hat{H}_{\rm col}$,
determines the in-band splitting and, possibly,
band mixing. The cubic $r_0$ term in
$\hat{H}$ induces the $\gamma$-dependence
in the energy surface~(\ref{E-surface})
and is essential for obtaining
the DF-like spectrum of Fig.~\ref{fig-spectra}.
For $r_0=0$, $\hat{H}$ reduces to
the Hamiltonian of the SO(6) dynamical symmetry (DS)
chain,
\ba
{\rm U(6)}\supset {\rm SO(6)}\supset
   {\rm SO(5)}\supset {\rm SO(3)} ~,
\label{SO6-DS}
   \ea   
with eigenvalues
$E = h_0(N-\sigma)(N+\sigma+4)
+ \rho_5\tau(\tau+3) +\rho_3L(L+1)$
and eigenstates
$\ket{N,\sigma,\tau,n_{\Delta},L}$
specified by quantum numbers of the
algebras in the chain ($n_{\Delta}$ is a multiplicity
label).
In this case,
the states in each SO(6) $\sigma$-band
form SO(5) $\tau$-multiplets.
For $\rho_3\!=\!0$, the rotational spectrum
in the $\sigma\!=\!N$ ground band
coincides with that of the WJ model
for a $\gamma$-unstable shape.
\begin{table*}[t]
\begin{center}
\caption{\label{Tabftau}
\protect\small
The factors $f_{\tau,n_{\Delta}}^{(L,K)}$,
Eq.~(\ref{Thetatb1}), for 
the states $\ket{N,LM;K}$, Eq.~(\ref{so6-decomp}),
with $(L,K)=(0,0),\,(2,2),\,(2,0),\,(3,2)$. 
The label $n_{\Delta}$ is not required, since 
these $L$-states are multiplicity-free.
$P_{n}(x)$ stands for a Legendre polynomial of
order $n$, which
satisfies
$P_{n}(0) = (-1)^{n/2}\,\frac{(n-1)!!}{n!!}$
for $n$ even and
$P_{n}(0) = 0$ for $n$ odd.
} 
\vspace{1mm}
\begin{tabular}{llll}
\hline
& & &\\[-3mm]
& $\qquad f_{\tau,n_{\Delta}}^{(L,K)}$ & \hspace{1.8cm}
SO(5) $\tau$-assignment & \\
\hline
& & &\\[-2mm]
$L=0,K=0\quad$ &
$\sqrt{3(2n+1)}P_{n}(0)\quad$ &
$\tau = 3n\;$,
$n$ even $(n=0,2,4,\ldots)\quad$ &
$\tau=0,6,12,\ldots$\\[4pt]
$L=2,K=2\quad$ &
$\sqrt{\frac{3(n+1)}{2}}P_{n}(0)\quad$ &
$\tau = 3n+1\;$,
$n$ even $(n=0,2,4,\ldots)\quad$ &
$\tau=1,7,13,\ldots$\\[2pt]
&
$\sqrt{\frac{3(n+1)}{2}}P_{n+1}(0)\quad$ &
$\tau = 3n+1\;$,
$n$ odd $(n=1,3,5,\ldots)\quad$ &
$\tau=5,11,17,\ldots$\\[4pt]
$L=2,K=0\quad$ &
$\sqrt{3(n+1)}P_{n+1}(0)\quad$ &
$\tau = 3n+1\;$,
$n$ odd $(n=1,3,5,\ldots)\quad$ &
$\tau=4,10,16,\ldots$\\[2pt]
&
$\sqrt{3(n+1)}P_{n}(0)\quad$ &
$\tau = 3n+2\;$,
$n$ even $(n=0,2,4,\ldots)\quad$ &
$\tau=2,8,14,\ldots$\\[4pt]
$L=3,K=2\quad$ &
$\sqrt{\frac{3(n+1)(2n+3)}{2(n+2)}}P_{n}(0)\quad$ &
$\tau = 3n+3\;$,
$n$ even $(n=0,2,4,\ldots)\quad$ &
$\tau=3,9,15,\ldots$\\
& & &\\[-3mm]
\hline
\end{tabular}
\end{center}
\end{table*}

The Hamiltonian of Eq.~(\ref{Hfull}) is invariant under
a discrete transformation
$(d^{\dag}_{\mu},\,d_{\mu})\rightarrow
(-d^{\dag}_{\mu},\,-d_{\mu})$,
hence commutes with the operator
$\hat{\cal R}_d = \exp(i\pi\,\hat{n}_d)$.
Consequently, all eigenstates
of $\hat{H}$
have well defined
$d$-parity~\cite{Rowe09,Thiamova12},
\ba
{\cal R}_d = \exp(i\pi\, n_d) = (-1)^{n_d} = (-1)^{\tau} ~.
\label{Rd}
\ea
The last equality follows from the
${\rm U(5)}\supset {\rm SO(5)}$
reduction which ensures that ($n_d-\tau$) is even.
The values of ${\cal R}_d$ coincide with those of
${\cal R}$-parity, ${\cal R} = (-1)^{n_{\gamma}+K/2}$,
in the collective model.
This can be inferred by examining the $(L,K)$-states
projected from
intrinsic states. For the ground ($g$),
$\beta$ and $\gamma$ bands, the intrinsic states are,
\bsub
\ba
&& \ket{g;N} =
\ket{\beq\!=\!1,\gaeq=\tfrac{\pi}{2};N} ~,
\label{int-g}
\\
\hspace{-1.5cm}
&& \ket{\beta;N} \approx
P^{\dag}_0\ket{g;N-2} ~,
\label{int-beta}
\\
\hspace{-1.5cm}
&& \ket{\gamma;N} \approx
W^{\dag}_0\ket{g;N-3} ~.
\label{int-gamma}
\ea
\label{int-gbetgam}
\esub
Here $\ket{\beta,\gamma ; N}$ is the ground band
intrinsic state, Eq.~(\ref{int-state}),
with the equilibrium deformations
and $(P^{\dag}_0,\,W^{\dag}_0)$ are the operators in
Eq.~(\ref{Hfull}).
By construction, $\ket{g,N}$ is an exact eigenstate of
$\hat{H}_{\rm int}$.  The intrinsic states
of excited bands, $\ket{\beta,N}$ and $\ket{\gamma,N}$,
become eigenstates only for large $N$. In that limit,
they involve a replacement of a condensate boson in
$\ket{g;N}$ by the normal modes,
$\ket{\beta; N}\to b^{\dag}_{\beta}\ket{g;N-1}$, 
$\ket{\gamma; N}\to b^{\dag}_{\gamma}\,\ket{g;N-1}$,
which identifies them as single-phonon bands.
$\ket{g;N}$ of Eq.~(\ref{int-g})
involves a sum of terms of the form
$(s^{\dag})^{N-k}(d^{\dag}_2)^{\ell}(d^{\dag}_{-2})^{k-\ell}\ket{0}$
with $n_d=k$ and $K/2=2\ell-k$ ($k,\,\ell$ integers).
It follows that the $(L,K)$ states projected from
$\ket{g;N}$ with $K/2$ even (odd),
involve U(5) basis states with $n_d$ even (odd)
hence, by Eq.~(\ref{Rd}), states of the ground band
have ${\cal R}_d=(-1)^{K/2}$.
Since $P^{\dag}_{0}$ ($W^{\dag}_{0}$) has even (odd)
$d$-parity, it follows from Eq.~(\ref{int-gbetgam})
that states of the $\beta$ and $\gamma$ bands have
${\cal R}_d=(-1)^{K/2}$  and ${\cal R}_d=(-1)^{1+K/2}$,
respectively.

The $r_0$ term of Eq.~(\ref{Hfull})
breaks the SO(6) dynamical symmetry,
Eq.~(\ref{SO6-DS}), and in general
mixes states with different $(\sigma,\tau)$.
A~notable exception are the states of the ground band
which retain a good SO(6) quantum number ($\sigma$).
This arises from the fact
that $P_0\ket{g;N}=0$, implying that
$\ket{g;N}$ of Eq.~(\ref{int-g})
has $\sigma\!=\!N$. $P^{\dag}_0$ is an SO(6)-scalar,
hence $\ket{\beta;N}$ of Eq.~(\ref{int-beta}) has
$\sigma\!=\!N-2$. $W^{\dag}_0$ is a $\sigma\!=\!3$
tensor with respect to SO(6), hence $\ket{\gamma;N}$
of Eq.~(\ref{int-gamma}),
contains a mixture of components with
$\sigma\!=\! N-2k\,(k=0,1,2,3)$,
obtained from the
product \mbox{$(\sigma\!=\!3)\otimes(\sigma\!=\!N-3)$}.
These SO(6) features are retained by the
SO(3) projection from intrinsic states.
Altogether, we find the following $d$-parity and
SO(6) properties for the rotational states
of the ground, $\beta$ and $\gamma$ bands,
\bsub
\ba
&&
\hspace*{-0.9cm}
\ket{g;N,L;K,{\cal R}_d\!=\!(-1)^{K/2}}\;
\hspace{0.36cm}
\sigma\!=\!N ~,\hspace{2cm}
\label{wf-g}\\
\hspace*{-3cm}
  &&
\hspace*{-0.9cm}
\ket{\beta;N,L;K,{\cal R}_d\!=\!(-1)^{K/2}}\;
\hspace{0.36cm}
\sigma\!=\! N-2 ~,
\label{wf-beta}\\
&&
\hspace*{-0.9cm}
\ket{\gamma;N,L;K,{\cal R}_d\!=\!(-1)^{1+K/2}}
\;\;
\sigma \!=\!
  N,N\!-\!2,N\!-\!4,
  \,N\!-\!6 ~.\;\;\qquad
\label{wf-gamma}
\ea
\label{wf-gbgam}
\esub
The above situation,
where some states preserve part of the symmetries
of the chain~(\ref{SO6-DS}), defines a
partial dynamical symmetry~\cite{lev11}.

The states
$\ket{N,LM;K}\!\equiv\!\ket{g;N,LM;K,
  {\cal R}_d\!=\!(-1)^{K/2}}$
of the triaxial ground band,
are obtained by standard
SO(3) projection from the intrinsic
state of Eq.~(\ref{int-g}),
\ba
\hspace{-0.65cm}
&&
\ket{N,LM;K} =
\sqrt{\tfrac{2^N}{N!(2L+1)}}
\left [\Gamma_{K,K}^{(N,L)}\right ]^{-1/2}
\hat{{\cal P}}^{(L)}_{M,K}\,
\ket{g;N}.\;
\label{state-LKb1}
\ea
Here
$\hat{{\cal P}}^{(L)}_{M,K}$
is the projection operator
and $K\!\geq\! 0$ even defines a set of independent
states. Each such state has well defined $d$-parity
${\cal R}_d\!=\!(-1)^{K/2}$.
The normalization $\Gamma_{K,K}^{(N,L)}$
is obtained from,
\ba
\hspace{-0.5cm}
\Gamma_{K_1,K_2}^{(N,L)} = \frac{1}{16\pi^2N!}
\int d\Omega
\left[1+ g(\Omega)\right]^N
d(L,K_1,K_2,\Omega) ~,\;
\label{integralb1}
\ea
where
$\Omega\equiv (\phi,\theta,\psi)$
denotes Euler angles, $d(L,K_1,K_2,\Omega)=$\\
$d^{(L)}_{K_1,K_2}(\theta)
\cos(K_1\phi\!+\!K_2\psi)
+ (-1)^{L}d^{(L)}_{K_1,-K_2}(\theta)
\cos(K_1\phi\!-\!K_2\psi)$ and
$g(\Omega) = \tfrac{1}{2}[
(1 + \cos^2\theta)\cos 2\phi\, \cos 2\psi
  -2\cos\theta\,\sin 2\phi\,\sin 2\psi]$.
Here
\mbox{\footnotesize $
d^{(L)}_{K_1,K_2}(\theta)$}
is the reduced rotation
matrix.
In general, the $(L,K)$-projected states are not
orthogonal.
An orthonormal set of states, retaining good $d$-parity,
can be obtained by a Gram-Schmidt procedure.

The finiteness of $N$ and the conserved $\sigma$
and ${\cal R}_d$  quantum numbers
govern the SO(5) structure of the
$(L,K)$-states~(\ref{state-LKb1}).
Their expansion in the SO(6) basis reads,
\ba
\hspace{-0.5cm}
\ket{N,LM;K} =
\sum_{\tau,n_{\Delta}}
\Theta^{(N,L,K)}_{\tau,n_{\Delta}}\,
\ket{N,\sigma\!=\!N,\tau,n_{\Delta},LM} ~,\;\;
\label{so6-decomp}
\ea
where the $\tau$ summation
covers the range $\tau=0,1,\ldots, N$, with values
of $(\tau,n_{\Delta})$ compatible with the
${\rm SO(5)}\supset {\rm SO(3)}$ reduction.
The conserved $d$-parity implies that states with
${\cal R}_d\!=\! (-1)^{K/2}\!=\!+1$ ($K\!=\!0,4,8,\ldots$) 
involve components with $\tau$ even,
while states with
${\cal R}_d\!=\! (-1)^{K/2}\!=\! -1$ ($K\!=\!2,6,10,\ldots$) 
involve components with $\tau$ odd.
The expansion coefficients are of the form,
\ba
  \Theta^{(N,L,K)}_{\tau,n_{\Delta}} =
  \left [\Gamma_{K,K}^{(N,L)}\right ]^{-1/2}
    \sqrt{\tfrac{2^{N+1}(N+1)!}
      {(N-\tau)!(N+\tau+3)!}}\,
  f_{\tau,n_{\Delta}}^{(L,K)} ~,\qquad
\label{Thetatb1}
  \ea
where $\Gamma_{K,K}^{(N,L)}$
is obtained from
Eq.~(\ref{integralb1}).
Analytic expressions of
$\Theta_{\tau,n_{\Delta}}^{(N,L,K)}$
for~states with $(L,K)=(0,0),(2,0),(2,2),(3,2)$
can be worked out. The corresponding
factors $f_{\tau,n_{\Delta}}^{(L,K)}$ and
permissible SO(5) $\tau$-components 
are listed in Table~\ref{Tabftau}.
For these states, $\Gamma_{K,K}^{(N,L)}$ can also be
obtained from the normalization condition
$\sum_{\tau,n_{\Delta}}[\Theta^{(N,L,K)}_{\tau,n_{\Delta}}]^2 = 1$.
This allows one to study the evolution
of structure with the size of the system.
Such a task was carried out in~\cite{Otsuka87}
for the $L=0$ ground state and is explored
here also for excited rotational states.
\begin{figure}[t]
\vspace{-0.9cm}
\hspace*{-0.85cm}
\includegraphics[width=1.22\linewidth]
                {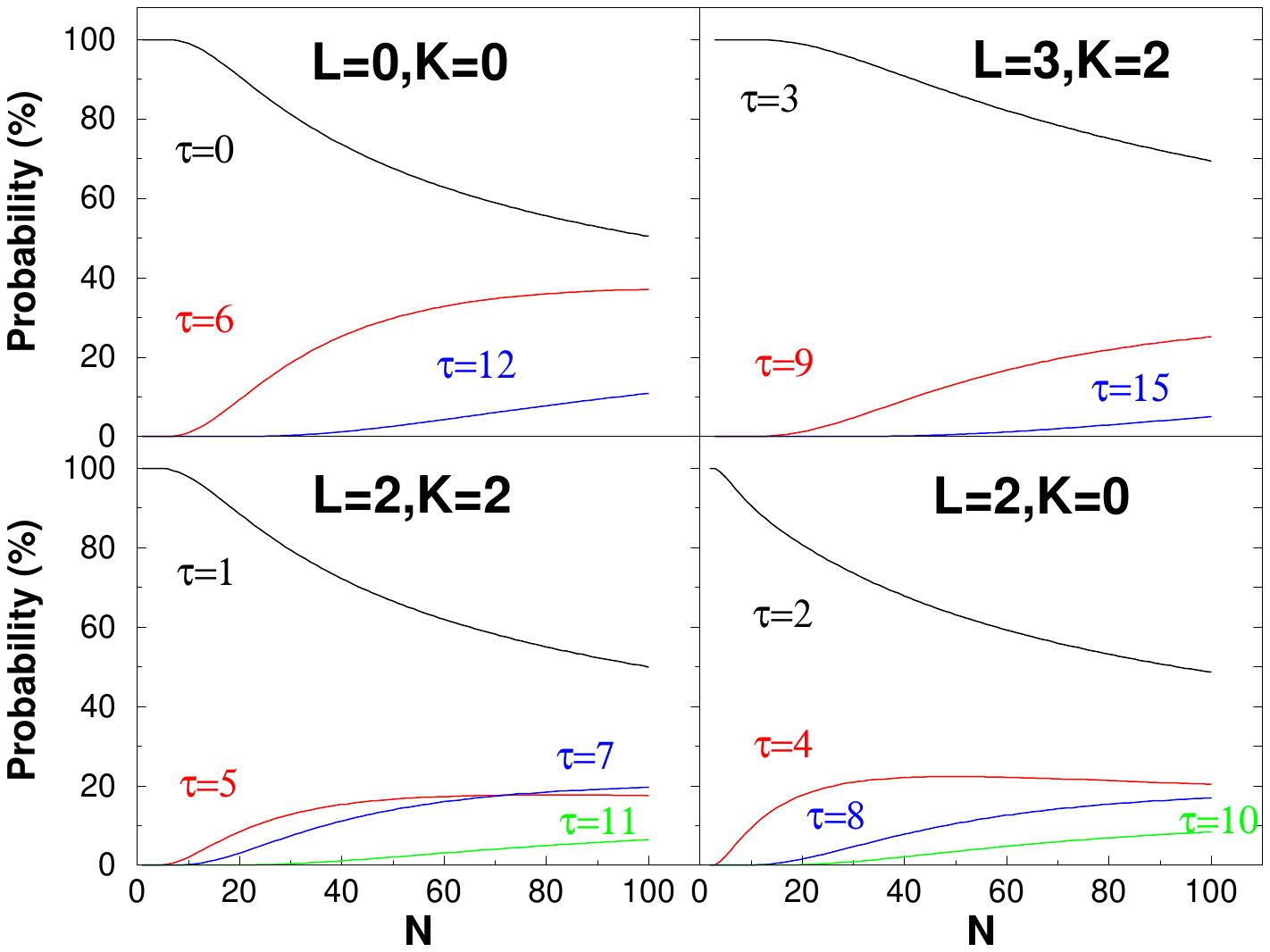}
  \caption{\label{fig-ProbLK}
  \small
  Probabilities
  $\left [\Theta^{(N,L,K)}_{\tau,n_{\Delta}}\right ]^2$ of
  dominant SO(5) $\tau$-components in the 
  states $\ket{N,L;K}$, Eq.~(\ref{so6-decomp}),
  with $(L,K)=(0,0),\,(2,2),(2,0),(3,2)$, as a function
  of $N$.
  }
\end{figure}

Figure~\ref{fig-ProbLK} shows the dominant
$\tau$-components
in the SO(5) decomposition of states,
as a function of~$N$.
When $N$ increases, one observes
a transition from $\gamma$-soft structure (SO(6)-DS)
to $\gamma$-rigid structure (DF, $\gamma=\pi/6$).
For small $N$, each of the $(L,K)$-states
displays a high degree of SO(5) purity and is
dominated by a single SO(6) basis state,
$\ket{N,\sigma\!=\!N,\tau,L}$, with the lowest allowed
value of $\tau$.
As $N$ increases, additional $\tau$ components
contribute, displaying a sharper change with $N$
for smaller values of $\tau$.
For large $N$, the $(L,K)$-states maintain
$\sigma\!=\!N$ as a good quantum number
but exhibit substantial $\tau$ mixing.
The SO(5) admixtures reach 50\% for
$(L,K)=(0,0),(2,0),(2,2)$
and 30\% for $(L,K)=(3,2)$.
The corresponding wave functions deviate considerably
from those of a $\gamma$-soft rotor (SO(6)-DS)
and approach properties of a rigid triaxial rotor
(DF, $\gamma=\pi/6$). This change in structure
will be reflected in the pattern of energies
and $E2$ transitions, to be discussed below.
\begin{figure}[t]
\vspace{-1.4cm}
\hspace*{-0.94cm}
\includegraphics[width=1.37\linewidth]
                {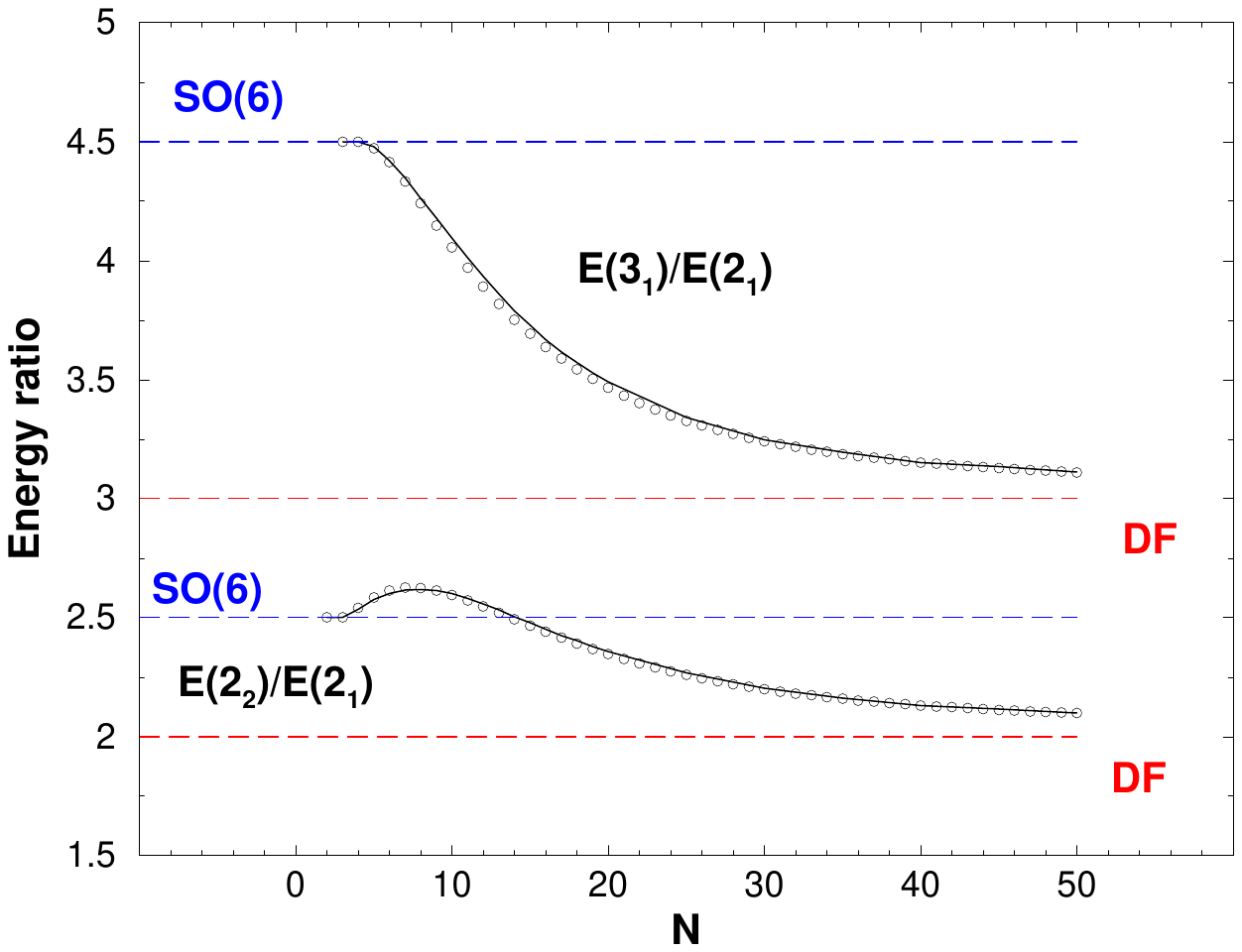}
\vspace{-0.9cm}
\caption{\label{fig-EneRatio}
\small
Energy ratios,
$\tfrac{E(2_2)}{E(2_1)}$ and
$\tfrac{E(3_1)}{E(2_1)}$,
as a function of~$N$,
calculated numerically
by diagonalizing
the Hamiltonian~(\ref{Hfull})
with parameters
$h_0\!=\!15,\,r_0\!=\!0.25,\,\rho_5\!=\!1,\,\rho_3\!=\!0$
(solid lines),
and analytically ($\circ$) using
Eq.~(\ref{Exc-gL}).
The ratios in the rigid triaxial (DF)
and $\gamma$-soft [SO(6)] limits are
marked by dashed~lines.
}
\end{figure}
\begin{figure}[!t]
\vspace{-0.9cm}
\hspace*{-0.65cm}
\begin{overpic}[width=1.32\linewidth]
  {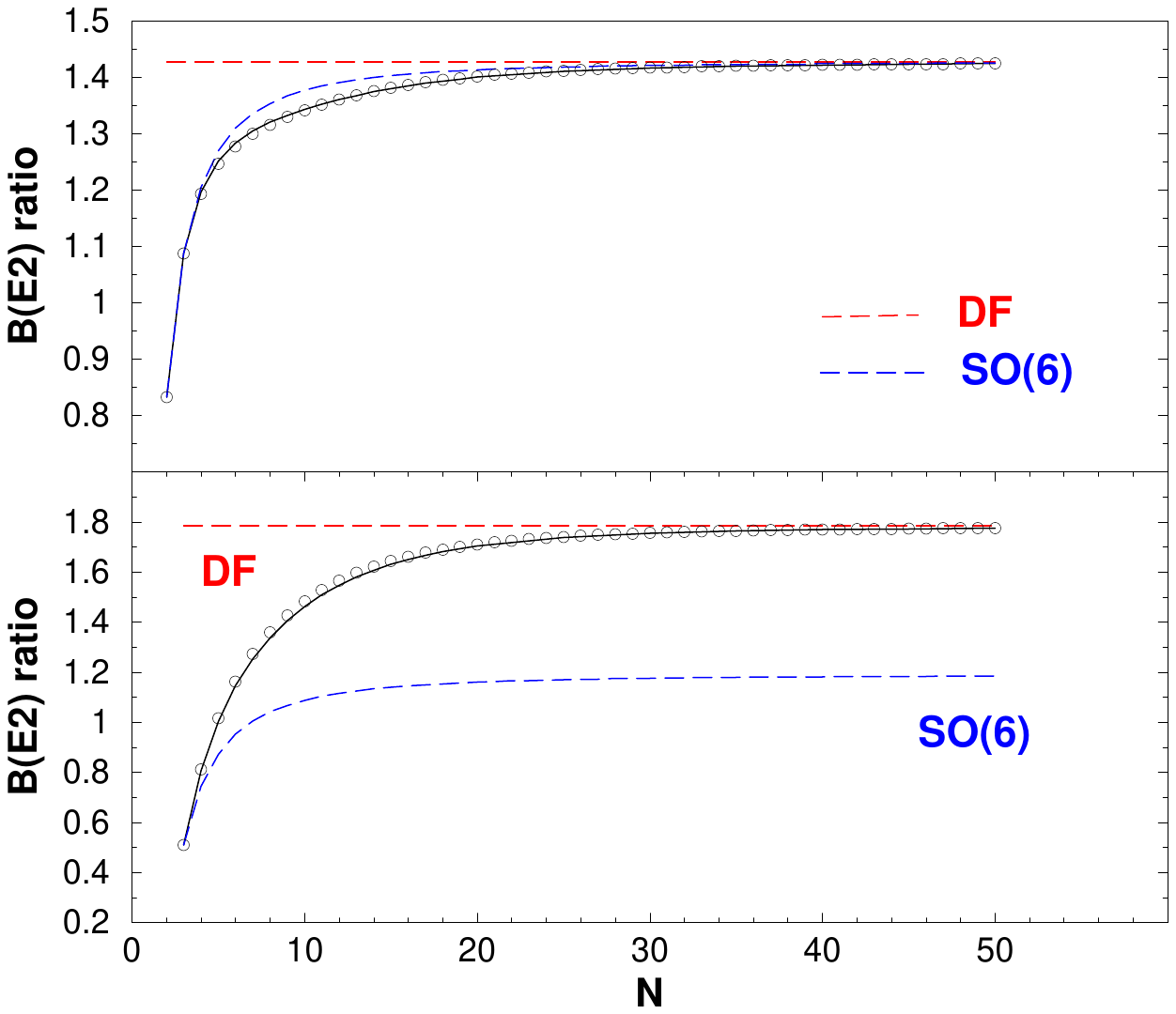}
\put(73,57) {(a)}
\put(73,29) {(b)}
\put (25,40) {\large
  $\frac{B(E2;2_2\to 2_1)}{B(E2;2_1\to 0_1)}$}
\put (25,12) {\large
  $\frac{B(E2;3_1\to 2_2)}{B(E2;2_1\to 0_1)}$}
\end{overpic}
\vspace{-0.45cm}
\caption{\label{fig-Be2Ratio}
  \small
  B(E2) ratios,
(a)~$\tfrac{B(E2;2_2\to 2_1)}{B(E2;2_1\to 0_1)}$ and
(b)~$\tfrac{B(E2;3_1\to 2_2)}{B(E2;2_1\to 0_1)}$,
as a function of $N$, calculated numerically (solid lines)
and analytically ($\circ$) using Eq.~(\ref{Be2}).
The ratios in the rigid triaxial (DF)
and $\gamma$-soft [SO(6)] limits are 
marked by dashed~lines.
}
\end{figure}

Having at hand explicit expressions for the wavefunctions
of states $\ket{N,LM;K}$ in the ground band,
Eq.~(\ref{state-LKb1}), one can derive
closed form expressions for various observables.
The expectation value of the Hamiltonian,
Eq~(\ref{Hfull}), provides an estimate for their
excitation energies,  
\ba
E_{N;L,K}\,
= \rho_3\, L(L\!+\!1)
-2(N\!+\!1)\rho_5
\,[
  S^{(N,L)}_{1,K} -S^{(N,L=0)}_{1,K=0}] ~.
\label{Exc-gL}
\ea
Here,
\ba
S^{(N,L)}_{1,K}\equiv
\bra{N,LM;K}
\hat{n}_s\ket{N,LM;K} =
\frac{\Gamma_{K,K}^{(N-1,L)}}
     {\Gamma_{K,K}^{(N,L)}} ~,\qquad
\label{SNL1K}
\ea
and $\Gamma_{K,K}^{(N,L)}$ are obtained from
Eq.~(\ref{integralb1}).

The quadrupole operator is taken to be,
\ba
\hat{T}(E2) = e_{b}(d^{\dag}s+s^{\dag}\tilde{d}) ~,
\label{Te2}
\ea
where $e_b$ is a boson effective charge
and $\tilde{d}_{m} = (-1)^{m}d_{m}$.
The transition probabilities (scaled by $e_{b}^2$)
for $E2$ transitions between the states
$\ket{g;N,L;K}$ of Eq.~(\ref{state-LKb1})
are found to be,
\ba
&&
\hspace{-1.1cm}
B(E2;g;L_i,K_i\to g;L_f,K_f) =
\frac{1}{2}\,
\frac{1}{
\Gamma_{K_f,K_f}^{(N,L_f)}\,
\Gamma_{K_i,K_i}^{(N,L_i)}}
\nonumber\\[2mm]
&&
\times
\left\{\,
C^{L_f,K_f}_{L_i,K_f-2;2,2}\,
\Gamma_{K_f-2,K_i}^{(N-1,L_i)}
+ C^{L_f,K_f}_{L_i,K_f+2;2,-2}\,
\Gamma_{K_f+2,K_i}^{(N-1,L_i)}
\right.
\nonumber\\[1mm]
&&
\hspace{0.5cm}
\left.
+\, C^{L_f,K_i-2}_{L_i,K_i;2,-2}\,
\Gamma_{K_f,K_i-2}^{(N-1,L_f)}
+ C^{L_f,K_i+2}_{L_i,K_i;2,2}\,
\Gamma_{K_f,K_i+2}^{(N-1,L_f)}\,\right\}^2.
\qquad
\label{Be2}
\ea
Here $C^{\ell_3m_3}_{\ell_1m_1;\ell_2m_2}$ denotes
a Clebsch-Gordan coefficient
and the factors $\Gamma_{K_1,K_2}^{(N,L)}$ are obtained
from Eq.~(\ref{integralb1}).
Henceforth, we examine
the evolution with $N$ of
these observables for a fixed Hamiltonian
with parameters listed in the caption of
Figure~\ref{fig-EneRatio}.
\begin{figure}[t]
\vspace{-1.1cm}
\hspace*{-1cm}
\begin{overpic}[width=1.3\linewidth]{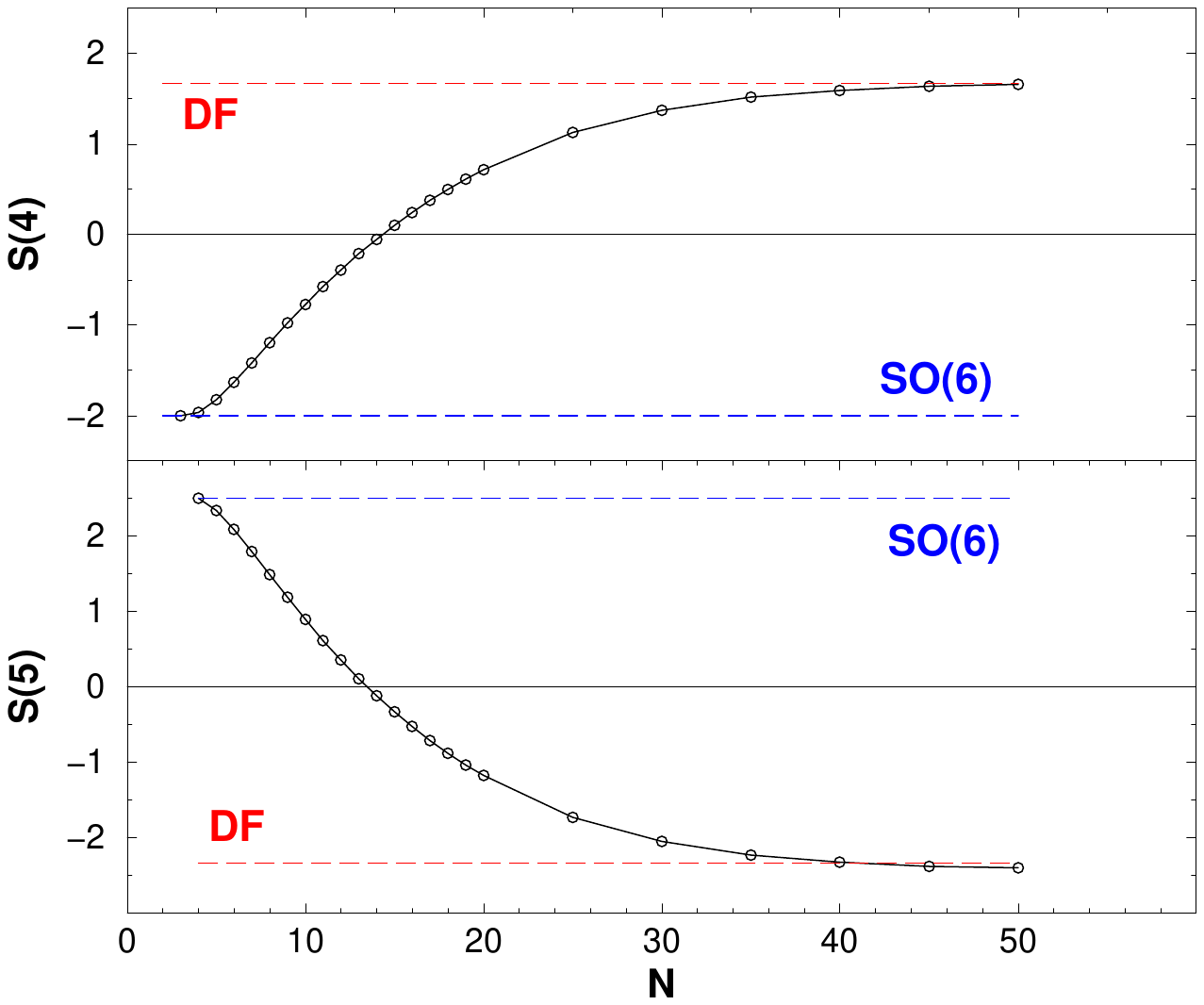}
\put(77,58) {(a)}
\put(77,30) {(b)}
\end{overpic}
\vspace{-0.9cm}
\caption{\label{fig-StagL4L5}
  \small
  Staggering index $S(L)$, Eq.~(\ref{stagg}), for
  states of the quasi-$\gamma$ band,
  (a)~$L=4_2$ and (b)~$L=5_1$, as a function of $N$.
  These states have, respectively, $K=2$ and $K=4$
  in the rigid triaxial (DF) limit
  with $S(4)\!=\!1.67,\,S(5)\!=\!-2.33$, and have
  $\tau=3$ and $\tau=4$
  in the $\gamma$-soft SO(6) limit with
  $S(4)\!=\!-2,\,S(5)\!=\!2.5$.
  }
\end{figure}
\begin{figure}[]
\vspace{-1.3cm}
\hspace*{-0.7cm}
  \includegraphics[width=1.32\linewidth]
                  {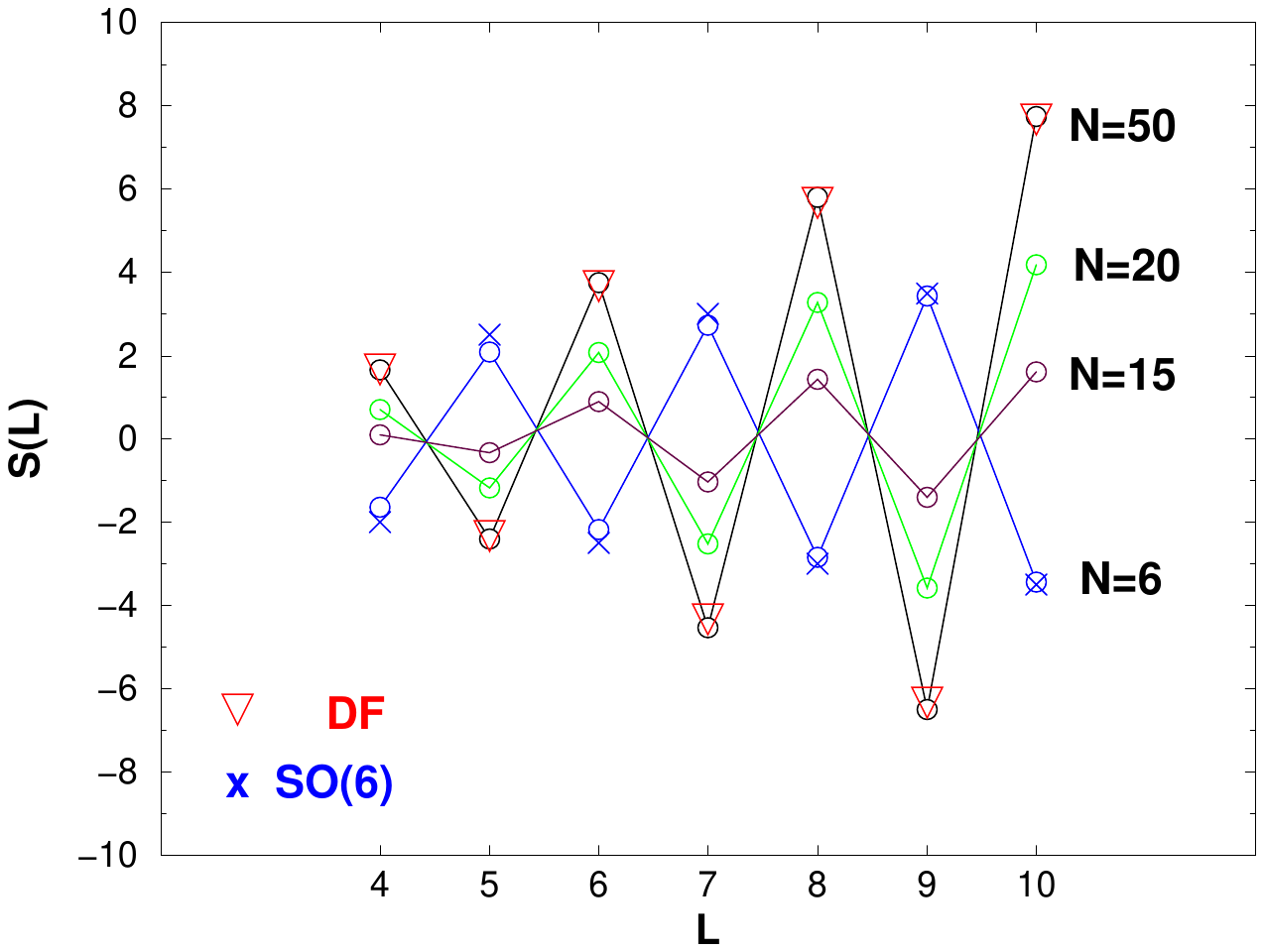}
\vspace{-0.9cm}
\caption{\label{fig-Stag}
  \small
  Staggering index $S(L)$, Eq.~(\ref{stagg}),
  for states in the quasi-$\gamma$ band as a function
  of $L$, for $N=6,\,15,\,20,\,50$.
  Note the smaller amplitude and change in the phase
  of $S(L)$ for $N=15$,
  in the transition from a $\gamma$-soft [SO(6)]
  pattern at small $N$
  (marked by \textcolor{blue}{$\times$})
    to a rigid-triaxial (DF) pattern at large $N$
    (marked by~\textcolor{red}{$\triangledown$}).
  }
\end{figure}

Consider the
energy ratios, $R_2 = \frac{E(2_2)}{E(2_1)},\,
R_3 = \frac{E(3_1)}{E(2_1)}$
and $B(E2)$ ratios,
$B_2 = \frac{B(E2;2_2\to 2_1)}{B(E2;2_1\to 0_1)},\,
B_3 = \frac{B(E2;3_1\to 2_2)}{B(E2;2_1\to 0_1)}$.
In the $\gamma$-rigid model (DF, $\gamma=\pi/6$)
with energy $E\propto [4L(L+4) -3K^2]$, the states
$0_1,\,2_1,\,2_2,\,3_1$ correspond, respectively, to
$(L,K) = (0,0),\,(2,2),\,(2,0),\,(3,2)$,
for which $R_2\!=\!2$, $R_3 \!=\! 3$,
$B_2 = \tfrac{10}{7}$ and $B_3=\tfrac{25}{14}$.
In the $\gamma$-soft model (SO(6)-DS),
with energy $E\propto \tau(\tau+3)$, these states
correspond to
$\tau = 0,\,1,\,2,\,3$, respectively, for which
$R_2\!=\!2.5$, $R_3 \!=\! 4.5$,
$B_2=\tfrac{10}{7}\tfrac{(N-1)(N+5)}{N(N+4)}$ and
$B_3=\tfrac{25}{21}\tfrac{(N-2)(N+6)}{N(N+4)}$.
Figures~\ref{fig-EneRatio} and~\ref{fig-Be2Ratio}
depict, respectively, the changes
in the energy ratios and
B(E2) ratios,
as a function of $N$.
As seen, these ratios
interpolate between the $\gamma$-soft values for
small~$N$, and the $\gamma$-rigid values for large~$N$.
The analytic expressions of Eqs.~(\ref{Exc-gL})
and~(\ref{Be2}), provide accurate
estimates to the calculated ratios.

Energy staggering in the quasi-$\gamma$ band
is a key signature for triaxiality.
The relevant quantity
is the staggering index~\cite{ZamCas91,McCtchan2007},
\ba
S(L) =
\frac{[E(L)\!-\!E(L-1)] - [E(L\!-\!1)\!-\!E(L\!-\!2)]}
{E(2^{+}_1)} ~.\qquad
\label{stagg}
\ea
In the rigid triaxial model (DF, $\gamma=\pi/6$),
the quasi-$\gamma$ band corresponds to
the sequence of rotational states
in the ground band
with ($L$~even, $K=L-2$) and 
($L$ odd, $K=L-1$)
with
$S(L) = \tfrac{1}{3}(3L-7) > 0$
for $L\!=\!4,6,8,\ldots$
and $S(L) \!=\! -\tfrac{1}{3}(3L-8) < 0$ for
$L\!=\!5,7,9,\ldots$.
In contrast, in the $\gamma$-soft model
(WJ and SO(6)-DS),
the quasi-$\gamma$ band corresponds to rotational
states with [$L$~even, $\tau=(L+2)/2$] and
[$L$ odd, $\tau=(L+3)/2$]
with
$S(L) = -\tfrac{1}{4}(L+4) < 0$
for $L\!=\!4,6,8,\ldots$
and
$S(L) = \tfrac{1}{4}(L+5) > 0$ for
$L\!=\!5,7,9,\ldots$.
The alternating signs and zig-zag pattern of
the staggering index signal deviations from
axial symmetry with
opposite phase for rigid and soft triaxiality.
In both cases, the magnitude of $S(L)$
increases with $L$.

Figure~\ref{fig-StagL4L5} portrays the evolution with
$N$ of the staggering index $S(L)$, Eq.~(\ref{stagg}),
for states in the quasi-$\gamma$
band with angular momenta $L=4,\,5$.
As shown, $S(L)$ obtain the
$\gamma$-soft SO(6) values
for small $N,$ and approach the rigid triaxial
values for large $N$.
In general, $S(L)$ for $L$ even, increases with~$N$ and
changes sign from negative to positive
values. In contrast, $S(L)$ for $L$ odd, decreases with
$N$ and changes sign from positive to
negative values. These opposite trends of $S(L)$ for
even and odd values of $L$, result in a change of phase
in the zig-zag patterns shown in Fig.~\ref{fig-Stag},
reflecting a transition from $\gamma$-soft to
$\gamma$-rigid structure, as $N$ increases.

The following observations are in place.
(i)~The transition point where $S(L)$ changes sign
for a given $L$, occurs at intermediate values of
$N\!\sim\!15$. This is consistent with the
fact that the transition from $\gamma$-soft to
$\gamma$-rigid structure necessitates substantial
SO(5) $\tau$-admixture in the wave functions of the
states $\ket{N,L;K}$ which, in turn, is limited
for small~$N$.
(ii)~Beyond the transition point,
the magnitude of $S(L)$ approaches the corresponding
rigid triaxial (DF) value
from below,
{\it i.e.}, $|S(L)|$ has smaller values.
The latter property highlights the role of finite-$N$
effects, and is consistent with the fact that
in candidate triaxial nuclei
(\textit{e.g.}, $^{76}$Ge, $^{112}$Ru, $^{170}$Er, $^{232}$Th),
the DF staggering indices
overshoot the empirical values.

\begin{figure}[t]
\vspace{-1.1cm}
\hspace*{-0.45cm}
\includegraphics[width=1.3\linewidth]{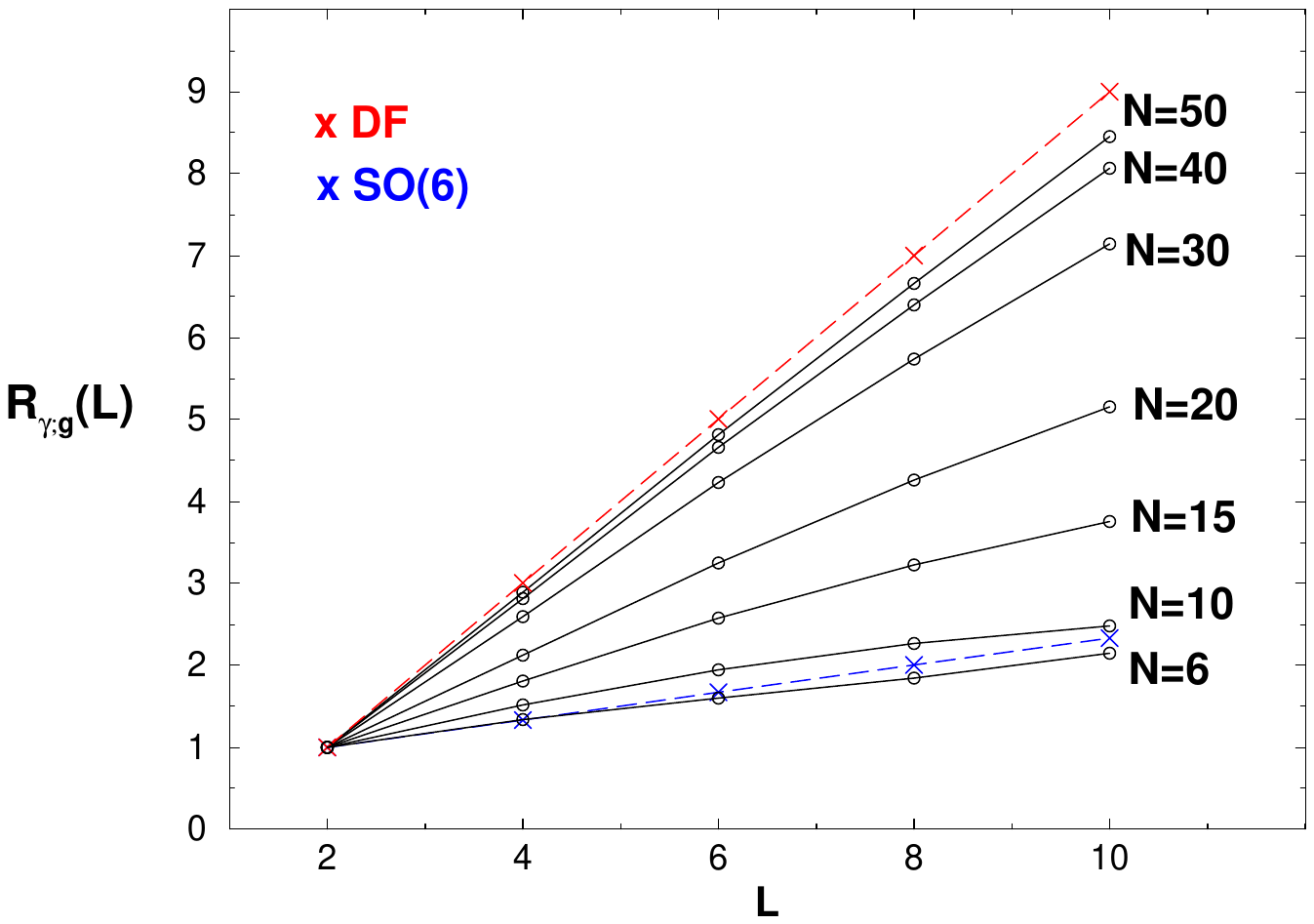}
\vspace{-1.1cm}
\caption{\label{fig-Rg-gam}
  \small
The ratios $R_{\gamma;g}(L)$ of Eq.~(\ref{Rg-gam})
  as a function of $L$, for $N=6,10,15,20,30,40,50$.
  The calculated $R_{\gamma;g}(L)$ curves
  approach the linear
  $\gamma$-soft curve at small $N$
  (blue dashed line), and
  the linear
  $\gamma$-rigid curve
  at large $N$ (red dashed line).
  }
\end{figure}

Another quantity that can distinguish between
$\gamma$-rigid and $\gamma$-soft behavior is the
ratio~\cite{Casten2020,Bonat21},
\ba
R_{\gamma;g}(L) =
\frac{E(L_{\gamma})-E(L_g)}{E(2_{\gamma})-E(2_g)}
\qquad L\geq 2\;{\rm even} ~. 
\label{Rg-gam}
\ea
It is a ratio of transition energies from the
quasi-$\gamma$ band to the quasi-ground band
for a given angular momentum $L$.
In the rigid triaxial model (DF, $\gamma=\pi/6$),
the quasi-ground band corresponds to the sequence of
yrast states with ($L\geq 0$ even, $K=L$)
and $R_{\gamma;g}(L) = \tfrac{1}{6}(L+4)$.
In the $\gamma$-soft model (WJ and SO(6)-DS), it
corresponds to the set of rotational states with
($L\geq 0$ even, $\tau=L/2$)
and $R_{\gamma;g}(L) = L-1$.
For both models the
$R_{\gamma;g}(L)$ ratios
are linear in $L$ but with different slopes.
Figure~\ref{fig-Rg-gam} depicts the ratio $R_{\gamma;g}(L)$
as a function of $L$ for various values of $N$.
At small $N$, the ratio is similar to the $\gamma$-soft
linear curve.
As $N$ increases,
$R_{\gamma;g}(L)$, increases slowly and beyond $N\sim 15$,
it increases more rapidly, gradually bending upwards
and approaches the $\gamma$-rigid
linear curve at large $N$.
\begin{table}[!t]
  \caption{\label{TableBe2}
\small
$B(E2)$ values
for transitions involving states
in the ground ($g$), $\gamma$ and $\beta$ bands.
The calculated
$E2$ rates (IBM) are in units of
$B(E2; g;2^{+}_1\to g;0^{+}_1) = 1$ and
employ the Hamiltonian of Eq.~(\ref{Hfull})
with $N = 25$
and parameters as in Fig.~\ref{fig-EneRatio}.
The calculated $E2$ rates (BM)
are taken from~\cite{Rowe09}
and employ the adiabatic Bohr model
Hamiltonian for a rigid triaxial shape with
$\gamma=\pi/6$.
The first BM values of the ground
$\gamma$ and $\beta$
transitions are chosen to match those of the
IBM calculation.
}
\begin{tabular}{lcc}
\hline
\\[-2pt]
Transition & IBM & BM ($\gamma=\frac{\pi}{6}$)
\\[2pt]
\hline
\\[-5pt]
$g;\, 2^{+}_1(K=2) \rightarrow g;\, 0^{+}_1(K=0)\quad$ &
1.00 & 1.00 \\
$g;\, 2^{+}_2(K=0) \rightarrow g;\, 2^{+}_1(K=2)\quad$ &
1.41 & 1.43 \\
$g;\, 3^{+}_1(K=2) \rightarrow g;\, 2^{+}_2(K=0)\quad$ &
1.74 & 1.79 \\
$g;\, 3^{+}_1(K=2) \rightarrow g;\, 4^{+}_1(K=4)\quad$ &
0.87 & 1.00 \\
$g;\, 4^{+}_1(K=4) \rightarrow g;\, 2^{+}_1(K=2)\quad$ &
1.40 & 1.39 \\
$g;\, 4^{+}_2(K=2) \rightarrow g;\, 2^{+}_2(K=0)\quad$ &
0.60 & 0.60 \\
$g;\, 4^{+}_2(K=2) \rightarrow g;\, 4^{+}_1(K=4)\quad$ &
0.36 & 0.27 \\
$g;\, 4^{+}_3(K=0) \rightarrow g;\, 2^{+}_1(K=2)\quad$ &
0.02 & 0.04 \\
$g;\, 4^{+}_3(K=0) \rightarrow g;\, 3^{+}_1(K=2)\quad$ &
0.63 & 0.56 \\
$g;\, 4^{+}_3(K=0) \rightarrow g;\, 4^{+}_2(K=2)\quad$ &
1.56 & 1.75 \\
$g;\, 5^{+}_1(K=4) \rightarrow g;\, 3^{+}_1(K=2)\quad$ &
0.94 & 0.95 \\
$g;\, 5^{+}_1(K=4) \rightarrow g;\, 4^{+}_2(K=2)\quad$ &
0.89 & 0.95 \\
$g;\, 5^{+}_1(K=4) \rightarrow g;\, 6^{+}_1(K=6)\quad$ &
0.64 & 0.71 \\
$g;\, 5^{+}_2(K=2) \rightarrow g;\, 4^{+}_1(K=4)\quad$ &
0.02 & 0.05 \\
$g;\, 5^{+}_2(K=2) \rightarrow g;\, 4^{+}_3(K=0)\quad$ &
1.52 & 1.59 \\
$g;\, 5^{+}_2(K=2) \rightarrow g;\, 5^{+}_1(K=4)\quad$ &
0.44 & 0.46 \\
& & \\
$\gamma;\, 2^{+}_3(K=2)\to \gamma;\, 0^{+}_2(K=0)\quad$ &
0.94 & 0.94 \\
$\gamma;\, 2^{+}_4(K=0)\to \gamma;\, 2^{+}_3(K=2)\quad$ &
1.32 & 1.34 \\[6pt]
$\gamma;\, 0^{+}_2(K=0)\to g;\, 2^{+}_2(K=0)\quad$ &
0.12 & 0.18 \\
$\gamma;\, 2^{+}_3(K=2)\to g;\, 2^{+}_1(K=2)\quad$ &
0.02 & 0.06 \\
$\gamma;\, 2^{+}_4(K=0)\to g;\, 2^{+}_2(K=0)\quad$ &
0.02 & 0.06 \\
$\gamma;\, 2^{+}_4(K=0)\to g;\, 0^{+}_1(K=0)\quad$ &
0.01 & 0.04 \\
 & & \\
$\beta;\, 2^{+}_5(K=2)\to \beta;\, 0^{+}_3(K=0)\quad$ &
0.63 & 0.63 \\
$\beta;\, 2^{+}_6(K=0)\to \beta;\, 2^{+}_5(K=2)\quad$ &
1.18 & 0.90 \\[6pt]
$\beta;\, 0^{+}_3(K=0)\to g;\, 2^{+}_1(K=2)\quad$ &
0.00 & 0.07 \\
$\beta;\, 2^{+}_5(K=2)\to g;\, 0^{+}_1(K=0)\quad$ &
0.00 & 0.01 \\
$\beta;\, 2^{+}_5(K=2)\to g;\, 2^{+}_2(K=0)\quad$ &
0.00 & 0.02 \\
$\beta;\, 2^{+}_6(K=0)\to g;\, 2^{+}_1(K=2)\quad$ &
0.00 & 0.02 \\[6pt]
$\beta;\, 0^{+}_3(K=0)\to \gamma;\, 2^{+}_4(K=0)\quad$ &
1.21 & - \\[4pt]
\hline
\end{tabular}
\end{table}

Both the $d$-parity and SO(6) properties of the states
and operators have important consequences for
$E2$ transitions and moments.
The states $\ket{i;L;K,{\cal R}_d}$
in each band ($i=g,\,\gamma,\,\beta$) have good
${\cal R}_d$ quantum number and the $E2$ operator of
Eq.~(\ref{Te2}) has negative d-parity.
This implies that
\ba
\hspace{-0.5cm}
B(E2;i;L;K,{\cal R}_d \to i';L';K',{\cal R}_{d}') = 0
\;\;
{\rm for}\;\;
{\cal R}_d\, {\cal R}_d' \!=\! +1 ~,
\label{E2-rules}
\ea
and all quadrupole moments vanish, 
$Q(i;L;K,{\cal R}_d ) = 0$.
Recalling the ${\cal R}_d$ assignments of the states
in Eq.~(\ref{wf-gbgam}),
we see that the above selection rules
translate to those of the adiabatic
triaxial model (DF, $\gamma=\pi/6$), namely,
that all $E2$ transitions
with $\Delta K=0$ are forbidden, except
for $\gamma\to g$ and $\beta\to \gamma$
($\Delta n_{\gamma}=\pm 1$).

$\hat{T}(E2)$
of Eq.~(\ref{Te2}) is a generator of
SO(6), hence cannot connect states in different
SO(6) irreducible representations ($\sigma$-irreps).
As mentioned in Eq.~(\ref{wf-gbgam}),
states of the ground, $\beta$ and $\gamma$ bands
have $\sigma=N$, $\sigma\approx N-2$, and
$\sigma \approx N,\,N-2,\,N-4,\,N-6$,
respectively. It follows that interband transitions
$\beta\nrightarrow g$ are forbidden,
while $\gamma\to g$ and $\beta\to\gamma$ transitions
are allowed.
One therefore expects the $\beta\to\gamma$
to be stronger than
the $\beta\to g$ transitions.
This situation is very different from that of the 
collective model where  
$\beta\to\gamma$ transitions involve a two-phonon
change, hence are weaker than
$\beta\to g$ transitions.
The allowed interband transitions obey
the $d$-parity selection rules of Eq.~(\ref{E2-rules}),
namely, $\gamma\to g$ and $\beta\to\gamma$ transitions
with $\Delta K=0$ ($\Delta K=2$)
are allowed (forbidden).
Figure~\ref{fig-spectra} demonstrates
the $d$-parity and SO(6) $\sigma$-
selection rules
for $E2$ transitions connecting $0^{+}$
and $2^{+}$ states. Importantly,
all excited $0^{+}$ states are vibrational
and their different decay pattern provides a signature
for identifying them as band-heads of
either $\gamma$ or $\beta$ bands.

The matrix elements of $\hat{T}(E2)$,
Eq.~(\ref{Te2}),
between intrinsic states~(\ref{int-gbetgam}),
reveal a hierarchy
for large $N$,
$g\to g,\,
\gamma\to\gamma,\,\beta\to\beta\,(\sim N)$,
$\gamma\to g\, (\sim\sqrt{N})$ and
$\beta\to\gamma\,(\sim 1)$ transitions.
The $\beta\to g$
intrinsic $E2$ matrix elements vanish,
consistent with a dominance of 
$\beta\to\gamma$ transitions
over $\beta\to g$.
These features are evident in Table~\ref{TableBe2}
which compares
the B(E2) values obtained from
the IBM Hamiltonian
of Eq.~(\ref{Hfull}) with $N\!=\!25$, and
from the adiabatic
Bohr model (BM) Hamiltonian~\cite{Rowe09} for
a rigid triaxial shape with
$\gamma=\pi/6$. The $E2$ transitions in the latter
model have adjustable scales which are fixed in the
manner described in the caption of
Table~\ref{TableBe2}.
As seen, for $N\!=\!25$ 
the calculated $E2$ rates for
$g\to g$, $\gamma\to\gamma$, $\gamma\to g$,
$\beta\to\beta$ are similar in both models and
the $\beta\to g$ transitions are weak.
In contrast, the $\beta\to\gamma$ transitions
are very different, being very strong in the IBM
calculation but forbidden, hence completely missing
in the BM calculation.

In summary, we have presented an
Hamiltonian appropriate for a rigid triaxial
roto-vibrator in a finite system. It obeys
a discrete symmetry ($d$-parity) and partial
SO(6) symmetry, with closed $N$-dependent expressions
for wave functions of states in the ground band
and related observables.
A detailed analysis of energy and B(E2) ratios
reveals the central role of finite-$N$ effects
and symmetries in governing the
nature of triaxiality and $E2$ properties.
For a fixed
Hamiltonian, the system is found to evolve
from a $\gamma$-soft
SO(6) structure, with high SO(5) purity for small $N$,
to a $\gamma$-rigid
structure, with substantial SO(5) admixtures
for large $N$.
For intermediate values of $N$,
the staggering index $S(L)$ and $R_{\gamma;g}(L)$
display the expected behavior but with smaller
amplitudes. $E2$ decays and selection rules
for states of the ground $\gamma$ and $\beta$ bands
are similar to those of the collective model.
A striking difference is the $\beta\to\gamma$ transitions
which dominate over $\beta\to g$ transitions.
The inherent simplicity, embedded symmetry structure
and analytic attributes turn
the proposed framework into
a suitable algebraic benchmark
for the study of $\gamma$-rigidity in nuclei.
It is of interest to extend the formalism,
\textit{e.g.}, by including
$d$-parity breaking terms in the Hamiltonian
and addressing
rigid triaxial shapes with a general $\gamma$-deformation
$(0 < \gamma <\tfrac{\pi}{3})$.

\end{document}